\documentclass[aps,pre,showpacs,twocolumn,amsmath,amssymb]{revtex4-1}

\usepackage{graphicx} 
\usepackage{color} 
\usepackage{amsmath}
\usepackage{enumerate}

\begin{document}

\title{Effect of strong wakes on waves in two-dimensional plasma crystals}

\author{T. B. R\"ocker}
\email{tbr@mpe.mpg.de}
\affiliation{Max Planck Institute for Extraterrestrial Physics, 85741 Garching, Germany}
\author{A. V. Ivlev}
\email{ivlev@mpe.mpg.de}
\affiliation{Max Planck Institute for Extraterrestrial Physics, 85741 Garching, Germany}
\author{S. K. Zhdanov}
\affiliation{Max Planck Institute for Extraterrestrial Physics, 85741 Garching, Germany}
\author{G. E. Morfill}
\affiliation{Max Planck Institute for Extraterrestrial Physics, 85741 Garching, Germany}


\begin{abstract}
We study effects of the particle-wake interactions on the dispersion and polarization of dust lattice wave modes in
two-dimensional plasma crystals. Most notably, the wake-induced coupling between the modes causes the branches to
``attract'' each other, and their polarizations become elliptical. Upon the mode hybridization the major axes of the
ellipses (remaining mutually orthogonal) rotate by $45^\circ$. To demonstrate importance of the obtained results for
experiments, we plot representative particle trajectories and spectral densities of the longitudinal and transverse waves --
these characteristics reveal distinct fingerprints of the mixed polarization. Furthermore, we show that at strong coupling
the hybrid mode is significantly shifted towards smaller wave numbers, away from the border of the first Brillouin zone
(where the hybrid mode is localized for a weak coupling).
\end{abstract}

\maketitle

\section{Introduction}

A complex (dusty) plasma consists of a weakly ionized gas and small solid particles (also referred to as ``dust'' or
``grains''), which are charged negatively due to absorption of the surrounding electrons and ions and typically carry some
thousands electron charges \cite{Shuklabook,Fortov05,Shukla03,Bonitz10,MorFor}. The name was chosen in analogy to complex
liquids, due to the fact that complex plasmas can be regarded as a new (plasma) state of soft matter \cite{Morfill09}. Under
certain conditions, strongly coupled complex plasmas can be treated as a {\it single-species} weakly-damped medium whose
interparticle interactions are determined by the surrounding electrons and ions \cite{Ivlev_book}.

\begin{figure}
	\includegraphics[width=\columnwidth,clip=]{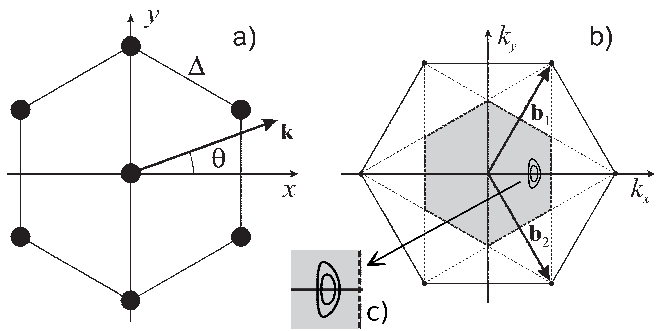}
	\caption{\label{refframe}
	(a) Elementary hexagonal lattice cell with the frame of reference. (b) The reciprocal lattice in ${\bf k}$-space with
    the basis vectors $|{\bf b}_{1}|=|{\bf b}_{2}|=\frac{4\pi}{\sqrt{3}}\Delta^{-1}$. Due to the lattice
	symmetry, it is sufficient to consider the wave vectors ${\bf k}$ at $0^{\circ}\leq\theta\leq 30^{\circ}$ and from within
	the first Brillouin zone (gray region enclosed by dashed lines), so that $|{\bf k}|\Delta\leq \frac{2\pi}{\sqrt{3}}$ for
	$\theta=0^{\circ}$ and $|{\bf k}|\Delta \leq\frac{4\pi}3$ for $\theta=30^{\circ}$.
	c) Enclosed regions where the particle-wake interactions cause the in-plane and out-of-plane wave modes to merge and
    form the hybrid mode (see also Fig.~\ref{flat}c). The smaller (larger) region corresponds to the effective dipole
    moment of wake $\tilde{d}_{\rm eff} = 0.7~(0.71)$ [Eq.~(\ref{dipole_eff})],
    the screening parameter is $\kappa = 1$, the normalized eigenfrequency of the vertical confinement is
    $\Omega_{\rm conf} = 5.66$.
}
\end{figure}

Experimentally, solid-state complex plasmas -- ``plasma crystals'' -- are produced by injecting monodisperse microparticles
in a low-pressure gas-discharge plasma \cite{Chu94,Thomas94}. In ground-based experiments performed in radio-frequency (rf) discharges, the
particles are normally levitated in the (pre)sheath above the powered horizontal rf electrode, where the balance of a
(highly inhomogeneous) electrostatic force and gravity produces a strong vertical confinement \cite{Ivlev_book,MorFor}. This provides excellent
conditions to create two-dimensional (2D) crystalline monolayers of particles with the hexagonal order, as shown in
Fig.~\ref{refframe}a.

The (pre)sheath electric field generates a strong vertical plasma flow. This produces a perturbed region downstream each
particle, which is called ``plasma wake'' (or simply ``wake'') \cite{Mela95,Vladi02}. The electrostatic potential of the wake has been calculated
analytically (by employing the linear-response formalism) and numerically (by using PIC-MCC
simulations) in very many papers, assuming different models for the ion velocity distribution and ion collisionality, see,
e.g., \cite{Hutchinson11,Vladi96,Ishihara97,Lampe2000,Ivlev05,Kompaneets07,Lampe12,Lapenta00}.

The wakes couple the horizontal and vertical motion of particles in the monolayer. The particle-wake interactions are
non-reciprocal \cite{Schweigert96}, and therefore under certain conditions the total kinetic energy of the ensemble is no longer
conserved: As was shown by Ivlev and Morfill \cite{Ivlev01}, when the horizontal and vertical motion are in resonance (which
occurs when the out-of-plane and longitudinal in-plane wave modes intersect) the {\it mode-coupling instability} sets in.
When the gas pressure is low enough and the instability cannot be suppressed by friction, microparticles acquire anomalously
high kinetic energy and the crystal melts. The mode-coupling instability and the associated melting has been investigated in
numerous theoretical \cite{Zhdanov09,Roecker12,Roecker12eff} and experimental works \cite{Ivlev03,Couedel09,Couedel10,Couedel11,Liu10err,Liu10corr}.

Self-consistent wake models provide better agreement with experimental results \cite{Roecker12,Roecker12eff}. Nevertheless,
many theoretical investigations of the mode-coupling instability are still based on a simple ``Yukawa/point-wake model''
\cite{Ivlev01,Zhdanov09}, since it makes the analysis of new qualitative effects associated with particle-wake interaction
more transparent. In this model, the wake is considered as a positive, point-like effective charge $q$ located at the
distance $\delta$ below each particle (of charge $Q$). Thus, the total interaction between two particles is a simple
superposition of the particle-particle and particle-wake interactions, both described by the (spherically-symmetric) Yukawa
potentials with effective screening length $\lambda$. For a given screening parameter $\kappa = \Delta/\lambda$ the effective dipole moment of wake,
\begin{equation}\label{dipole_eff}
\tilde{d}_{\rm eff}=\frac{\tilde{q} \tilde{\delta}}{1-\tilde{q}},
\end{equation}
(normalized by $Q \Delta$ and expressed in terms of the dimensionless wake charge $\tilde{q}=q/Q$ and distance $\tilde{\delta} = \delta/\Delta$),
determines the growth rate of the mode-coupling instability
\cite{Couedel11}.

The effective dipole moment is related to parameters of the ambient plasma and can be fairly large for typical experimental
conditions \cite{Roecker12eff}. On the other hand, the theoretical analysis of the mode coupling performed so far was
primarily focused on the limit of small $\tilde d_{\rm eff}$ \cite{Ivlev01,Zhdanov09,Couedel11}. The question of how the
wave modes are modified at large (but still experimentally accessible) values of $\tilde d_{\rm eff}$ remains open.
Furthermore, while uncoupled in-plane and out-of-plane wave modes (corresponding to purely horizontal and vertical motion,
respectively) are linearly polarized, the polarization of waves with a \emph{strong} wake-induced coupling (when $\tilde
d_{\rm eff}$ cannot be considered as a small parameter) is a fundamental question which has never been systematically
studied.

In this paper we investigate the regime of \emph{strong} wake-induced coupling between wave modes in 2D plasma crystals. In
Sec.~\ref{gen} we compare the weak and strong coupling regimes and point out their essential differences. The polarization
of separate and hybrid wave modes are discussed in Sec.~\ref{polin}. To demonstrate importance of the obtained results for
experiments and demonstrate distinct fingerprints of the mixed polarization upon the strong coupling, in
Sec.~\ref{observables} we plot representative particle trajectories and spectral densities of the longitudinal and
transverse waves, and calculate the shift of the hybridization point.

We normalize the wave number $k$ by the lattice constant $\Delta$, and all frequencies -- by the dust-lattice frequency
scale,
\begin{equation}\label{normal}
      \Omega_{\rm DL} = \sqrt{\frac{(1-\tilde{q})Q^2}{M \lambda^3}},
\end{equation}
viz.,
\begin{equation*}
	 k \Delta \rightarrow k, \qquad \omega/\Omega_{\rm DL},\Omega/\Omega_{\rm DL} \rightarrow \omega,\Omega,
\end{equation*}
where $M$ is the particle mass \cite{Ivlev01,Zhdanov09}.

\section{Weak and strong mode coupling}\label{gen}

The wake-induced coupling of the in-plane and out-of-plane modes is most efficient for the propagation direction $\theta =
0^{\circ}$ \cite{Zhdanov09,Couedel11}, when ${\bf k} = k {\bf \hat{e}}_x$ (see Fig.~\ref{refframe}a). In this case, the
in-plane transverse (acoustic ``shear'') mode becomes exactly decoupled. Therefore, below we assume $\theta = 0^{\circ}$ and
study the coupling between the remaining two modes \cite{Ivlev01,Zhdanov09}.

\begin{figure}
	\includegraphics[width=0.9\columnwidth]{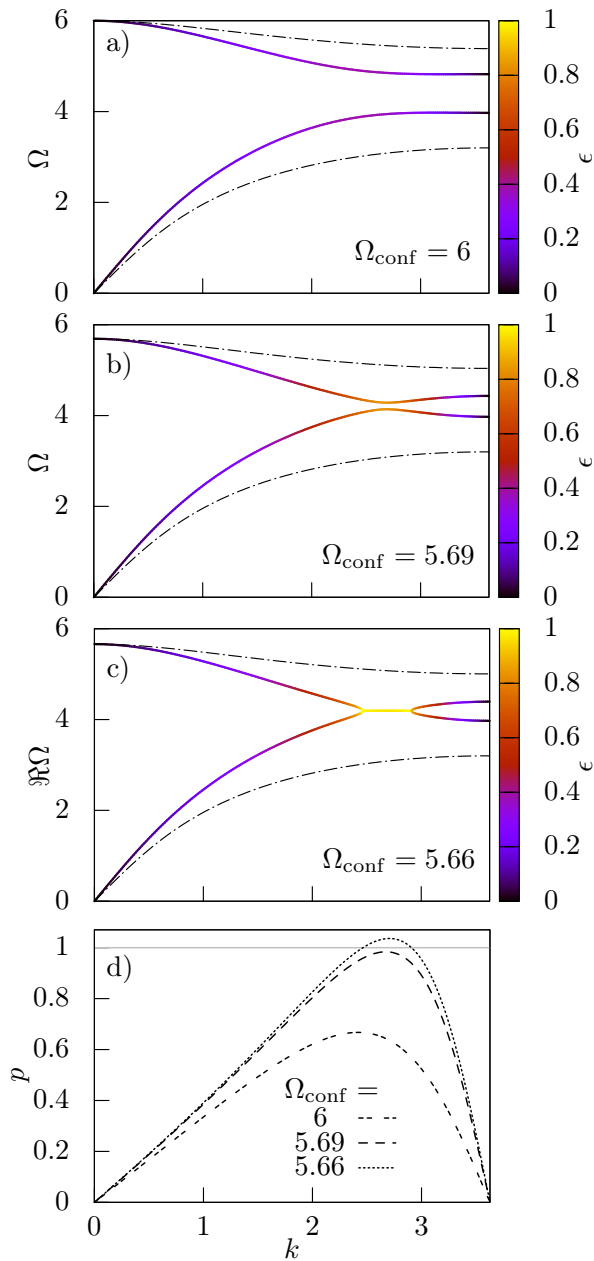}
	\caption{\label{flat}
	 Effect of the wake-induced coupling on the wave modes. (a-c) The colored lines show the (undamped) dispersion relations
    of the lower and upper modes, $\Omega_{\rm lo,up}(k)$ [Eq.~(\ref{eigenvalues1})], the dash-dotted lines represent the
    respective wake-free results, $\Omega_{\rm h,v}(k)$ [Eq.~(\ref{entr2D}) with $\tilde q=0$]. The curves are
    obtained for three different values of the vertical confinement frequency $\Omega_{\rm conf}$. In (a) the modes are
    attracted to each other, in (b) the gap between the modes is getting constricted, and in (c) they
    merge and form the hybrid mode (only the real part is shown). The color-coding shows the circularity of the mode
    polarization, $\epsilon$, which is determined by the reduced coupling parameter [Eqs~(\ref{eps1}) and
    (\ref{eps2}) for the separate and hybrid modes, respectively]. (d) The coupling parameter $p(k)$ [Eq.~(\ref{p})] for
    panels (a-c), the modes merge at $p = 1$. The results are for $\kappa = 1$, the dimensionless wake charge $\tilde{q}
    = 0.7$, the wake length $\tilde{\delta}=0.3$ ($\tilde{d}_{\rm eff}=0.7$).
}
\end{figure}

The exact dispersion relations $\omega(k)$ of the coupled in-plane and out-of-plane wave modes for $\theta = 0^{\circ}$ are
determined by the following equation \cite{Ivlev01,Zhdanov09,Couedel11}:
\begin{equation}\label{disp}
	\left[\omega(\omega+i\nu)-\Omega^2_{\rm h}\right]\left[\omega(\omega+i\nu)-\Omega^2_{\rm v}\right] + \sigma^2 = 0,
\end{equation}
where $\Omega_{\rm h,v}(k)$ are the eigenfrequencies of the {\it horizontal} (in-plane) and {\it vertical} (out-of-plane)
modes in the absence of coupling, $\sigma(k)>0$ is the coupling term, and $\nu$ is the damping rate due to neutral gas
friction \cite{Epstein24}. Expressions for $\Omega^2_{\rm h,v}$ and $\sigma$ are given in Appendix~\ref{elements} (wake-free
results are obtained by putting the wake charge $\tilde q$ equal to zero). Adopting the notation $\Omega^2 =
\omega(\omega+i\nu)$, Eq.~(\ref{disp}) can be written in a form of the eigenvalue problem \cite{Zhdanov09,Couedel11},
\begin{equation}\label{eigen}
	\det(\textsf{D}_{k}-\Omega^2 \textsf{I}) = 0, \qquad
	\textsf{D}_{k} =
	\left( \begin{array}{cc}
    \Omega^2_{\rm h} & i\sigma \\
    i\sigma & \Omega^2_{\rm v}
  \end{array}
\right),
\end{equation}
where $\textsf{I}$ denotes the unit matrix and $\textsf{D}_{k}$ is the non-Hermitian dynamical matrix (note that
$\textsf{D}_{k}^*=\textsf{D}_{-k}$). The squared eigenfrequencies of the {\it lower} and {\it upper} modes,
\begin{equation}\label{eigenvalues1}
	\Omega_{\rm lo,up}^2=\frac{\Omega^2_{\rm v}+\Omega^2_{\rm h}}{2}\mp\frac{\Omega^2_{\rm v}-\Omega^2_{\rm h}}{2}\sqrt{1-p^2},
\end{equation}
are determined by the {\it reduced coupling parameter} $p(k)$,
\begin{equation}\label{p}
	p = \frac{2 \sigma}{\Omega^2_{\rm v}-\Omega^2_{\rm h}}.
\end{equation}
When $p \rightarrow 0$ the lower and upper modes are reduced to the horizontal and vertical modes, respectively, i.e.,
$\Omega_{\rm lo,up}(k)\rightarrow \Omega_{\rm h,v}(k)$.

The eigenfrequencies of the upper and lower modes are real and different for $p<1$ -- in this regime we shall call the modes
{\it separate}. For $p> 1$ the eigenfrequencies become complex conjugate (i.e., with equal real parts), and therefore the
resulting lower and upper modes are referred to as \emph{hybrid}; in this case, $\Omega_{\rm lo,up}$ in
Eq.~(\ref{eigenvalues1}) has to be replaced with $\Omega^{{\rm (hyb)}}_{\rm lo,up}$. The eigenfrequencies of the lower
and upper hybrid modes have negative and positive imaginary parts, respectively, so that the former is decaying and the
latter is growing. The hybrid modes emerge at the {\it critical point} $p=1$, which occurs at the critical wave number $k =
k_{\rm cr}$ when the critical confinement $\Omega_{\rm conf} = \Omega_{\rm cr}$ is reached.

Thus, the wake-induced mode coupling can be considered as a {\it non-equilibrium} second-order phase transition, where $p$
plays the role of the control parameter, while the {\it imaginary part} of the hybrid eigenfrequency,
$\Im\Omega^{\rm(hyb)}$, is the proper order parameter.

The lower and upper modes near the onset of hybridization are illustrated in Fig.~\ref{flat}a-c for three characteristic
values of the vertical confinement frequency, $\Omega_{\rm conf}$ (for simplicity we consider undamped waves). The
corresponding coupling parameters $p(k,\Omega_{\rm conf})$ are plotted in Fig.~\ref{flat}d, so that $\Omega_{\rm conf}$
plays the role of the control parameter. In Fig.~\ref{flat}a the branches $\Omega_{\rm lo}(k)$ and $\Omega_{\rm up}(k)$ are
noticeably attracted to each other near $k \simeq 2.7$, where the maximum of $p(k)$ is reached; although $p_{\rm
max}\simeq0.67$ is still well below unity, the deviation from the wake-free results $\Omega_{\rm h,v}(k)$ is already quite
significant. In Fig.~\ref{flat}b, where $p_{\rm max}\simeq0.97$, one can see that the gap between the modes becomes
constricted. In Fig.~\ref{flat}c, where $\Omega_{\rm conf}>\Omega_{\rm cr}$, the modes merge and form the hybrid branch in
the range of $k$ where $p(k)\geq1$ [$\Re\Omega^{\rm (hyb)}(k)$ is shown]. In the ${\bf k}$-plane, the hybrid modes are
located in an enclosed region, as illustrated in Figs~\ref{refframe}b,c.

The magnitude of the coupling term $\sigma$ is determined by the effective dipole moment of the wake [see, e.g.,
Eq.~(\ref{entr2D})]. For the example presented in Fig.~\ref{flat} we choose $\tilde{d}_{\rm eff}=0.7$, which is rather large
but quite realistic for experiments (see, e.g. \cite{Ivlev03,Couedel10,Liu10corr}). As we can see, in this case the critical point and hence the hybrid
branch are significantly shifted away from the border of the Brioullin zone, where the hybrid mode is localized in the limit
of weak coupling [since the uncoupled branches $\Omega_{\rm h,v}(k)$ are monotonic].

The ``attraction effect'' between the modes seen in Fig.~\ref{flat}a-c has not been carefully studied before, since
theoretical analysis of the mode coupling performed so far was primarily focused on the limit of small $\sigma$
\cite{Ivlev01,Zhdanov09,Couedel11}. As we can see from Eq.~(\ref{p}), the critical point in this limit is reached when the
uncoupled branches practically cross. From Fig.~\ref{flat} it is evident that for finite $\sigma$ the merging occurs well
before the crossing (which now corresponds to $p \rightarrow \infty$).

In what follows we shall always consider a weak neutral damping, viz., $\nu \ll \Omega_{\rm lo,up}$, which is the typical
situation for experiments \cite{Couedel09,Couedel10,Ivlev03}. In this case the dispersion relations $\omega_{\rm lo,up}(k)$
are obtained by simply adding $-\frac12 \nu$ to the imaginary part of $\Omega_{\rm lo,up}(k)$, i.e.,
\begin{equation}\label{fric}
	\omega_{\rm lo,up}(k) \simeq \Omega_{\rm lo,up}(k)-i\frac{\nu}2.
\end{equation}
On the other hand, we shall assume the damping to be sufficiently strong to suppress the mode-coupling instability, viz.,
$\nu> 2|\Im\Omega^{\rm(hyb)}(k)|$. This ensures that the kinetic temperature of particles can reach a steady-state level
which is low enough to keep the crystalline order, i.e., the hybrid mode can still be observed in experiments.

\section{Polarization of wave modes}\label{polin}

From the eigenvectors $\boldsymbol{\zeta}_{\alpha}$ of the dynamical matrix (where $\alpha =$~lo, up) we obtain the wave
modes ${\bf w}_{\alpha}$. The separate modes are
\begin{equation}\label{wave}
	{\bf w}_{\alpha} = \boldsymbol{\zeta}_{\alpha}e^{ikx-i\omega_{\alpha}(k)t} + c.c.,
\end{equation}
where $c.c.$ denotes the complex conjugate, and the hybrid modes ${\bf w}_{\alpha}^{\rm(hyb)}$ are similarly obtained from
$\boldsymbol{\zeta}^{\rm (hyb)}_{\alpha}$.

Polarization of the modes can be conveniently characterized by the ``real phases'',
\begin{equation}
    \Phi_{\alpha} \equiv kx -\Re\Omega_{\alpha} t, \nonumber
\end{equation}
and the ``circularity'' $\epsilon$: $0$ and $1$ stand for the linear and circular polarizations, respectively,
$0<\epsilon<1$ are for elliptical polarization. For the separated and hybridized regimes the circularity is given by certain
functions of the reduced coupling parameter $p$ (see below).

\subsection{Separate modes ($p<1$)}\label{sep}

\begin{figure}
	\includegraphics[width=\columnwidth]{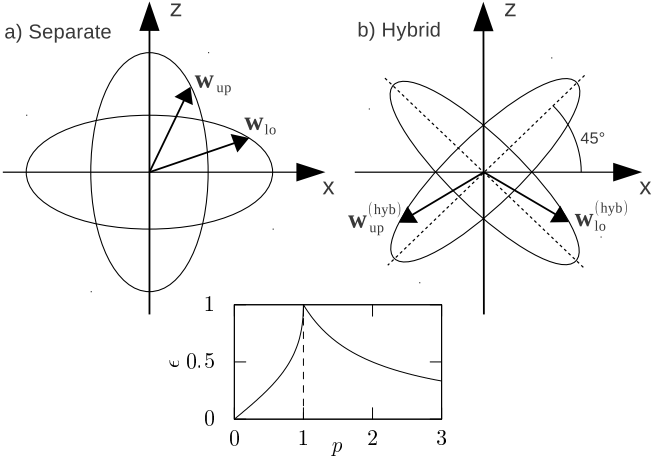}
	\caption{\label{elpol}
	Lissajous ellipses of the lower/upper {\it separate} modes ${\bf w}_{\rm lo,up}$ [Eq. (\ref{wavevec})] and lower/upper
{\it hybrid} modes ${\bf w}^{\rm (hyb)}_{\rm lo,up}$ [Eq. (\ref{hybres2})]. For the separate modes the major axes of the
ellipses are along the coordinate axes, for the hybrid modes the axes remain mutually orthogonal, but are rotated by
$45^{\circ}$. The shape of the ellipses is determined by the circularity $\epsilon$: For $\epsilon = 0$ the ellipses
degenerate into lines, for $\epsilon=1$ they form a circle (the illustration is for $\epsilon = 0.5$). The inset below shows
the dependence of $\epsilon$ on the reduced coupling parameter $p$, which is determined by Eqs~(\ref{eps1}) and (\ref{eps2})
for the separate and hybrid modes, respectively.
	}
\end{figure}

Uncoupled ($p \rightarrow 0$) horizontal and vertical wave modes are linearly polarized ($\epsilon \rightarrow 0$) and the
corresponding dynamical matrix is purely real (Hermitian) \cite{Zhdanov09}. Hence, the uncoupled eigenvectors are mutually orthogonal and real. For finite $p$ we obtain
\begin{equation}\label{ev}
	\boldsymbol \zeta_{\rm lo} = \frac{\left(1,\; -i\epsilon \right)}{\sqrt{1+\epsilon^2}}, \quad
    \boldsymbol \zeta_{\rm up} = \frac{\left(i \epsilon,\; 1 \right)}{\sqrt{1+\epsilon^2}},
\end{equation}
so that $|{\boldsymbol \zeta}_{\alpha}|^2 = 1$ and the circularity $\epsilon$ is the following function of $p$:
\begin{equation}\label{eps1}
	\epsilon = p/(1+\sqrt{1-p^2}).
\end{equation}
Thus, the lower and upper modes expressed via the corresponding wave phases,
\begin{equation}\label{wavevec}
    \begin{array}{r}
	{\displaystyle{\bf w}_{\rm lo}=\frac{2e^{-\frac{1}{2}\nu t}}{{\sqrt{1+\epsilon^2}}}\left(\cos\Phi_{\rm lo},\;\epsilon \sin\Phi_{\rm lo}\right),}
    \vspace{0.2cm} \\
	{\displaystyle{\bf w}_{\rm up}=\frac{2e^{-\frac{1}{2}\nu t}}{{\sqrt{1+\epsilon^2}}}\left(-\epsilon \sin\Phi_{\rm up},\;\cos\Phi_{\rm up}\right),}			
    \end{array}	
\end{equation}
are parametric functions of the two ellipses illustrated in Fig.~\ref{elpol}a.

Figure~\ref{flat} shows that for $k\ll1$ as well as for $k$ near the border of the first Brillouin zone the coupling
parameter $p$ is very small and therefore both ${\bf w}_{\rm lo}$ and ${\bf w}_{\rm up}$ are linearly polarized ($\epsilon
\rightarrow 0$). For $k$ where $p\simeq p_{\rm max}$ the circularity becomes significant and the polarization is essentially
elliptical. At the points where the upper and lower separate branches merge and form the hybrid branch (Fig.~\ref{flat}c)
the modes are circularly polarized ($\epsilon = 1$).

\subsection{Hybrid modes ($p\geq1$)}\label{hybrid}

For $p = 1$ the modes are degenerate. Hence, the eigenvectors for the separate modes [Eq.~(\ref{ev})] and hybrid modes
should coincide at this point, which yields
\begin{equation}\label{ev2}
    \begin{array}{r}
	{\displaystyle{\boldsymbol \zeta}^{\rm (hyb)}_{\rm lo}=\frac1{\sqrt{2}}\left(1,\;-i\epsilon-\sqrt{1-\epsilon^2}\right),}
    \vspace{0.2cm} \\
    {\displaystyle{\boldsymbol \zeta}^{\rm (hyb)}_{\rm up}=\frac1{\sqrt{2}}\left(i\epsilon+\sqrt{1-\epsilon^2},\;1\right).}
    \end{array}
\end{equation}
The lower and upper hybrid modes,
\begin{equation}\label{hybres2}
\begin{array}{r}
	{\displaystyle{\bf w}^{\rm (hyb)}_{\rm lo} = \sqrt2 e^{-\frac{1}{2}(\gamma + \nu)t}\; \{\cos\Phi_{\rm lo},\; \sin(\Phi_{\rm lo}-\Phi_0)\},}
    \vspace{0.2cm} \\
    {\displaystyle{\bf w}^{\rm (hyb)}_{\rm up} = \sqrt2 e^{\frac{1}{2}(\gamma - \nu)t}\; \{-\sin(\Phi_{\rm up}-\Phi_0),\; \cos\Phi_{\rm up}\},}
\end{array}
\end{equation}
are parametric functions of the two ellipses shown in Fig.~\ref{elpol}b. Here, $\gamma(k)\equiv 2|\Im\Omega^{\rm(hyb)}(k)|$
is the damping (growth) rate of the lower (upper) hybrid mode, the circularity $\epsilon$ and phase shift $\Phi_0$ are given
by
\begin{equation}\label{eps2}
      \epsilon = \cos \Phi_0 = p^{-1}.
\end{equation}

We see that the major polarization axes of the hybrid modes (remaining mutually orthogonal) are rotated by $45^\circ$ with
respect to the axes of separate modes. The rotation occurs discontinuously at $p=1$ (where the degenerate modes are
circularly polarized). Unlike the separate modes, the hybrid mode circularity decreases with $p$ (see inset in Fig. \ref{elpol}), i.e., the hybrid modes are
always elliptically polarized between the merging points (see Fig.~\ref{flat}c). Furthermore, upon the crossing of the horizontal and vertical branches,
$\Omega_{\rm v}^2-\Omega_{\rm h}^2\to0$, the coupling $p$ diverges at the crossing point and the polarization of hybrid
modes becomes linear near the intersection.

So far, reliable experimental results on the onset of the mode-coupling instability have been only obtained for a slightly
over-critical coupling. The reason for that is rather simple -- upon a ``deep quenching'' the crystal cannot be observed
long enough (to deduce sufficiently accurate dispersion relations) before it is completely destroyed by the instability.
Such conditions (e.g., of Fig.~\ref{flat}c) yield slightly distorted circular polarization ($0.95 \lesssim \epsilon \leq 1$)
along the whole hybrid branch.

\section{Features of strong coupling: Examples}\label{observables}

The strong mode coupling introduces several characteristic features which are expected to be revealed upon analysis of
corresponding experiments. Below we discuss some of the most interesting phenomena characterizing the individual particle
motion as well as the wave modes near the hybridization point.

\subsection{Particle trajectories}\label{part}

\begin{figure}
	\includegraphics[width=0.8\columnwidth]{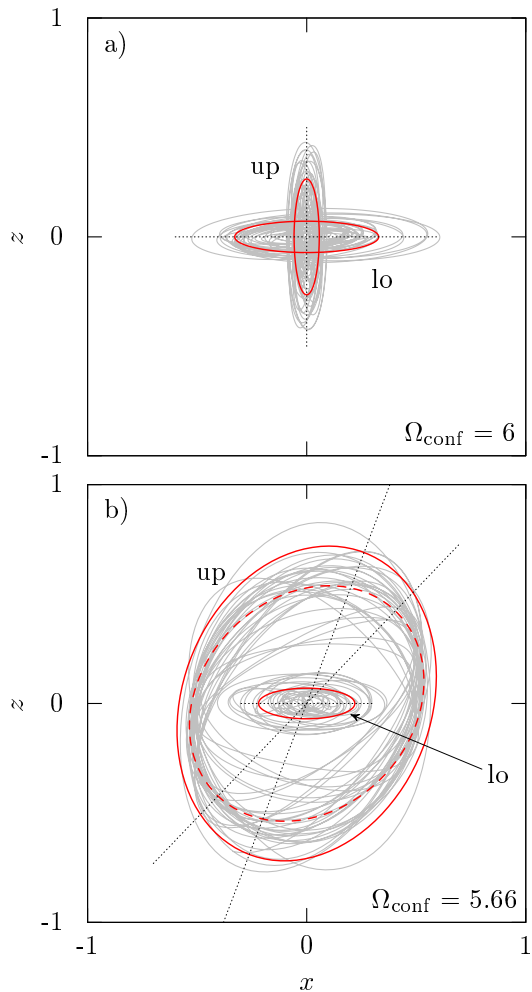}
	\caption{\label{tra}
        Characteristic particle trajectories representing the lower and upper wave modes for the time interval of
        $50~\Omega_{\rm DL}^{-1}$. All parameters of (a) and (b) correspond to Fig.~\ref{flat}a and c, respectively, $x$ and
        $z$ are in arbitrary units. The solid-line ellipses depict ``orbitals'' (straight dotted lines indicate their major
        axes) inside which the particles mostly reside. In the separated regime (a), the major axes are along the coordinate
        axes. In the hybridized regime (b), the major axis for the (decaying) lower mode is practically horizontal, while
        for (growing) upper mode it is rotated by $\simeq70^{\circ}$. The dashed-line ellipse shows the contribution
        from the upper hybrid branch only.}
\end{figure}

The individual particle trajectories are calculated by applying the inverse Fourier transform (over $k$) to the wave modes,
Eqs~(\ref{wavevec}) and (\ref{hybres2}). Thus, each trajectory can be viewed as the weighted ``superposition'' of the
Lissajous ellipses (at a given $k$), with the weights determined by the magnitudes of the corresponding eigenvectors. The
gray lines in Fig.~\ref{tra} illustrate the resulting steady-state trajectories (when friction is on average balanced by the
thermal noise, see Sec.~\ref{spectra} for details) for the separated (a) and hybridized (b) regimes. The trajectories are
shown for the time interval of $50~\Omega_{\rm DL}^{-1}$, which corresponds to $\sim3$~s under typical experimental
conditions \cite{Couedel11}. We also draw ``orbitals'' (solid-line ellipses), to indicate the areas mostly covered by the
trajectories (i.e., where particles reside most of the time). Due to stronger vertical confinement in the separated regime
(Fig.~\ref{tra}a), the vertical extent of the upper-mode orbital is somewhat smaller than the horizontal size of the
lower-mode orbital.


In the hybridized regime illustrated in Fig.~\ref{tra}b, the trajectories are determined by the superposition of (weighted)
contributions from the separate and hybrid branches (see Fig.~\ref{flat}c). As one can conclude from  Fig.~\ref{elpol}, this
yields a ``superposition'' of ellipses oriented along the coordinate axes and those rotated by $45^\circ$. The hybrid
branches are formed within a relatively small region of the Brillouin zone, yet the amplitude of the upper branch can be
significantly increased, while for the lower branch it is reduced [see Eq.~(\ref{hybres2})]. All this governs the size and
orientation of the corresponding orbitals: The lower orbital is practically horizontal and its size is similar to that in
Fig.~\ref{tra}a, since the amplitude for the lower hybrid branch is too small and hence the separate branch provides the
major contribution here. Contrary, the upper orbital is large and rotated by an angle notably smaller than $90^\circ$ (but
larger than $45^\circ$), since the major contribution in this case is from the enhanced hybrid branch (the corresponding
orbital is indicated by the dashed-line ellipse).

\subsection{Mode spectral density}\label{spectra}

\begin{figure*}
	\includegraphics[width=1.7\columnwidth]{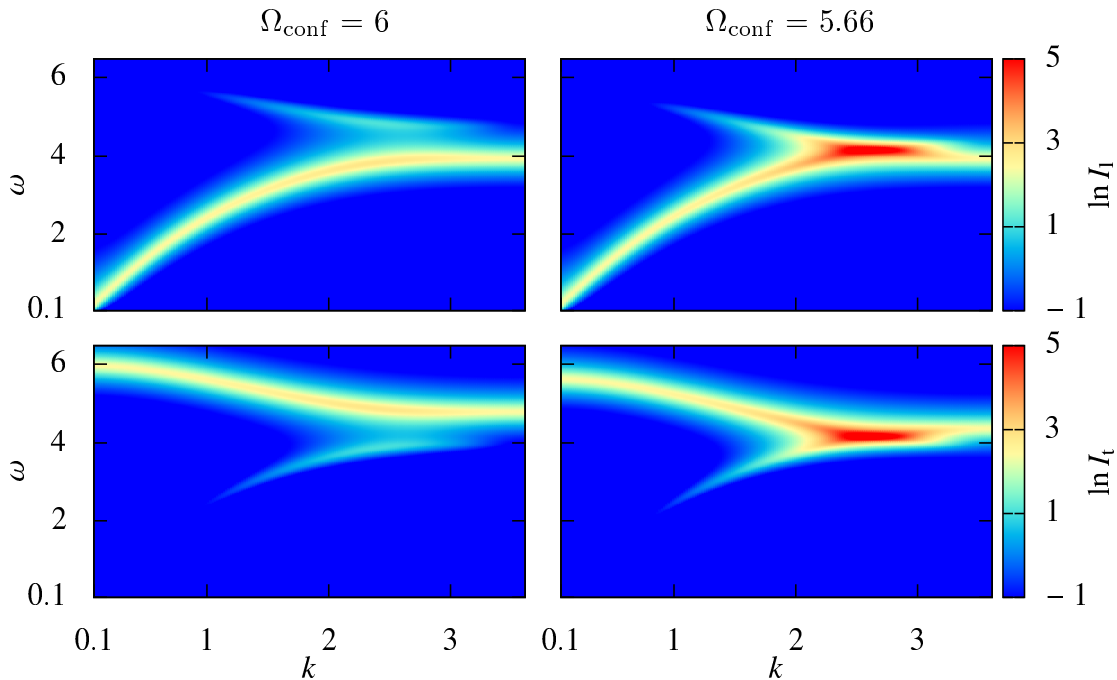}
	\caption{\label{itens}
		Longitudinal (upper panel) and transverse (lower panel) mode spectral densities $I_{\rm l,t}(k,\omega)$. Figures on
the left show the results calculated from Eq.~(\ref{inten1}) for the separate modes, on the right the results from
Eq.~(\ref{inten2}) for the hybrid modes are presented (all parameters correspond to Figs~\ref{flat}a and \ref{flat}c,
respectively). In the separated regime, the lower mode (longitudinal without coupling) has a transverse component, the upper
mode (transverse without coupling) has a longitudinal component. The longitudinal and transverse densities of the hybrid
modes are equal, which follows from the orientation of their polarization ellipses (see Fig.~\ref{elpol}b). The shown
example is for the normalized damping rate $\nu = 0.3$, the spectral density is in units of $I_0$. }	
\end{figure*}

Experimentally, fluctuation wave spectra (dispersion relations) are extracted from the observation of the particle motion
which is triggered by thermal noise and damped by neutral gas friction \cite{Nunomura02,Couedel09,Couedel10}.

The scenario can be mimicked by the Langevin equation, where the stochastic excitation force on each particle is
uncorrelated and described by the Gaussian white noise \cite{Schwabl,Zhdanov03,VanKampenBook}. We write the force in the
form $M{\bf a}(t)$, where the elements $a_j$ of the stochastic acceleration are defined by the first two stochastic moments,
$\langle a_j(t)\rangle=0$ and $\langle a_j(t)a_k(t') \rangle = (2\nu T/M) \delta_{jk}\delta(t-t')$ (as it follows from the
fluctuation-dissipation theorem). Here, $\delta_{jk}$ is the Kronecker delta function ($j,k=x,y,z$) and $T$ is the relevant
kinetic temperature (determined by the temperature of neutral gas and mechanisms of individual dust ``heating'' operating in
a plasma, e.g., due to charge variations \cite{Vaulina99,Ivlev00,Nunomura99,Pustylnik06}).

The perturbations for the separate modes read
\begin{equation}\label{mod}
	\tilde{\boldsymbol{\zeta}} = \sum_{\alpha=\rm lo, up} \frac{\left({\bf a}^{(k\omega)}\cdot\boldsymbol{\zeta}_{\alpha}\right)
\boldsymbol{\zeta}_{\alpha}}{[\Omega_{\alpha}^2(k)-\omega(\omega+i\nu)]\zeta_{\alpha}^2},
\end{equation}
where $\zeta_{\alpha}^2\equiv \boldsymbol{\zeta}_{\alpha}\cdot\boldsymbol{\zeta}_{\alpha}$ and ${\bf a}^{(k\omega)}$ is the
Fourier transform of the stochastic acceleration over {\it all} particles; for the hybrid modes, the superscript ``(hyb)''
has to be added.

For a monolayer comprised of $N$ particles, the total kinetic energy is
\begin{equation}
 K = \frac12M\sum_{j=1}^N \left\langle v_j^2 \right\rangle \equiv N\int d{\bf k} \int d\omega\; I({\bf k},\omega).
\end{equation}
Here $I({\bf k},\omega)=I_{\rm l}({\bf k},\omega)+I_{\rm t}({\bf k},\omega)$ is the spectral density of thermal
fluctuations, which consists of the longitudinal and transverse components (with respect to ${\bf k} = k {\bf \hat{e}}_x$,
the ${\bf k}$-integration is over the first Brillouin zone). By representing $K$ via the Fourier-transformed particle
velocities with the components $v^{(k\omega)}_{\rm l,t} = -i\omega \tilde{\zeta}_{x,z}$, we obtain the spectral density
components expressed in terms of the horizontal and vertical eigenfrequencies:
\begin{eqnarray}
	I_{\rm l,t}(k,\omega) = I_0\omega^2\frac{\sigma^2+|A_{\rm v,h}|^2}{|\sigma^2+A_{\rm v}A_{\rm h}|^2}
     \label{inten1}
\end{eqnarray}
where $A_{\rm h,v}(k,\omega)\equiv\Omega_{\rm h,v}^2(k)-\omega(\omega+i\nu)$. The spectral density scale $I_0=\sqrt3
\Delta^2 \nu T/\left(16\pi^3\Omega^2_{\rm DL}\right)$ has the dimensionality of J$\cdot$cm$^2\cdot$s (here $\nu$ is not
normalized!).

Remarkably, $I_{\rm l}(k,\omega)$ of the separate modes shows not only a bright longitudinal, but also a weak transverse
component, and vice versa for $I_{\rm t}(k,\omega)$ (see left panel of Fig.~\ref{itens}). When the coupling is
intermediate-to-strong ($0.3 \lesssim p \lesssim 0.95$, see Fig.~\ref{flat}d) the ``weak'' components become fairly bright.
For a small coupling ($p \ll1$) Eq.~(\ref{inten1}) is reduced to $I_{\rm l,t}(k,\omega)\simeq I_0\omega^2 \big|{\Omega^2_{\rm
h,v}(k) -\omega(\omega+i\nu)}\big|^{-2}$ and the weak components disappear. Qualitative interpretation of these results is
straightforward: The mode polarization changes between linear and elliptical, and therefore we find the ``weak'' components
varying between zero and their peak brightness. The latter is smaller than the density of the ``strong'' components, since
the modes are still separated.

From the orientation of polarization ellipses of the hybrid modes (Fig.~\ref{elpol}, $p>1$), it is evident that their
longitudinal and transverse spectral densities are equal along the hybrid branch (see right panel of Fig.~\ref{itens}).
These densities are
\begin{eqnarray}
I_{\rm l}^{\rm(hyb)}(k) = I_{\rm t}^{\rm(hyb)}(k)\nonumber\\[.2cm]
	=I_0\frac{\nu^2+(p^2+1)B}{[\nu^2-(p^2-1)B]^2},
\label{inten2}
\end{eqnarray}
where $B(k)=\frac12(\Omega_{\rm v}^2-\Omega_{\rm h}^2)^2(\Omega_{\rm v}^2+\Omega_{\rm h}^2)^{-1}$. In accordance with
Eq.~(\ref{hybres2}), the decay rate of the lower hybrid mode
is added to the neutral damping rate $\nu$, while the growth rate of the upper mode
is subtracted. Therefore, the relative contribution of the lower hybrid mode to the total spectral density in
Eq.~(\ref{inten2}) is relatively small. The density naturally diverges if the damping threshold of the mode-coupling
instability is reached [at the maximum of $p(k)$].

Figure~\ref{itens} indicates that the spectral density distribution along the hybrid branch can be slightly asymmetric (with
respect to the center). This is because the functions $p(k)$ and $B(k)$ are slightly asymmetric as well (see
Fig.~\ref{flat}c,d). This feature might be important for the analysis of high-resolution experimental spectra
\cite{Couedel09,Couedel10,Couedel11,Liu10err,Liu10corr}.

\subsection{Shift of the critical point}\label{nleff}

\begin{figure}
	\includegraphics[width=0.85\columnwidth]{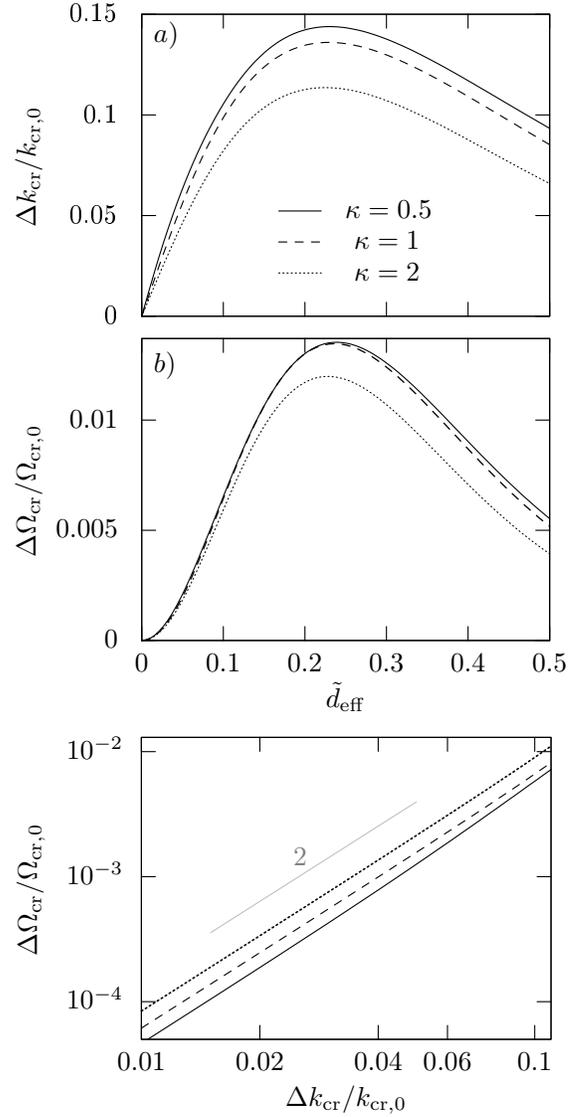} \\
	\caption{\label{kcrfig}
    The relative shift (a) of the critical wave number, $\Delta k_{\rm cr}/k_{\rm cr,0}$, and (b) of the critical confinement,
    $\Delta \Omega_{\rm cr}/\Omega_{\rm cr,0}$, as a function of the effective dipole moment of wake $\tilde{d}_{\rm eff}$.
    (c) Relation between the shifts, showing that it is very close to $\Delta \Omega_{\rm cr}\propto (\Delta k_{\rm cr})^2$. The
    results are for $\tilde{q} = 0.4$ and three different values of $\kappa$.}
\end{figure}

In the limit of {\it weak} mode coupling, the hybridization sets in upon the crossing of the uncoupled branches $\Omega_{\rm
h}(k)$ and $\Omega_{\rm v}(k)$. The critical point in this case is located at the border of the first Brillouin zone,
$k_{\rm cr,0} = \frac{2 \pi}{\sqrt{3}}$ (see Fig.~\ref{refframe}b), and the corresponding critical confinement is
$\Omega_{\rm cr,0}$. As one can see from Fig.~\ref{flat}a-c, in the regime of \emph{strong} coupling the critical point is
shifted towards smaller $k$ and occurs at higher confinement. Hence, we can write
\begin{equation}\label{shifts}
	k_{\rm cr} = k_{\rm cr,0} - \Delta k_{\rm cr}, \qquad \Omega_{\rm cr} = \Omega_{\rm cr,0} + \Delta \Omega_{\rm cr}.
\end{equation}
In Appendix~\ref{shifteq} we derive universal expressions for $k_{\rm cr}$ and $\Omega_{\rm cr}/\Omega_{\rm cr,0}$ which are
functions of the wake parameters $\tilde q$ and $\tilde\delta$ (for a given $\kappa$). Let us analyze the obtained results.

The functional dependence of $\Delta k_{\rm cr}$ on the wake parameters is governed by the competition of two effects: (i)
Increase of the effective dipole moment $\tilde{d}_{\rm eff}$ makes the ``mode attraction'' stronger, which causes $\Delta
k_{\rm cr}$ to \emph{increase}. (ii) Increase of the wake length $\tilde{\delta}$ weakens the screened wake-particle
interaction, which forces $\Delta k_{\rm cr}$ to \emph{decrease}. Figure~\ref{kcrfig}a illustrates these effects. We fixed
the wake charge $\tilde{q}$ and vary the wake length $\tilde{\delta}$ (and, hence, the dipole moment): When $\tilde{\delta}$
is sufficiently small ($\lesssim 0.2$), the effect (i) obviously dominates, resulting in an almost linear increase of
$\Delta k_{\rm cr}$ with the dipole moment; the maximum is reached at $\tilde{d}_{\rm eff} \simeq0.22$. For larger
$\tilde{\delta}$ ($\gtrsim 0.3$), the effect (ii) becomes more important and $\Delta k_{\rm cr}$ starts to decrease. When
the wake length is so large that the particle-wake interactions are strongly screened ($\kappa\tilde{\delta} \gtrsim 2-3$),
$\Delta k_{\rm cr}$ rapidly falls off and the modes become weakly coupled again.

It is noteworthy that the $k$-shift of the critical point can be really significant (up to $\simeq15\%$), making the effect
experimentally observable.

The dependence of $\Delta \Omega_{\rm cr}$ on $\tilde{d}_{\rm eff}$ shown in Fig.~\ref{kcrfig}b follows essentially the same
trend, since the two effects mentioned above equally work for the shift of the critical confinement. However, the magnitude
of $\Delta \Omega_{\rm cr}$ turns out to be much smaller than that of $\Delta k_{\rm cr}$ (${\Delta \Omega_{\rm
cr}}/{\Omega_{\rm cr,0}} \lesssim 2\%$) and therefore is practically negligible. To a very good accuracy, $\Delta
\Omega_{\rm cr}/{\Omega_{\rm cr,0}}$ is a parabolic function of $\Delta k_{\rm cr}$, as shown in Fig.~\ref{kcrfig}c.

Note that these results can be also obtained from a simple 1D string model with next-neighbor interactions \cite{Ivlev01}.
The 1D model provides compact analytical expressions for $\Delta k_{\rm cr}$ and ${\Delta \Omega_{\rm cr}}/{\Omega_{\rm
cr,0}}$ [see Eqs~(\ref{cr1D}) and (\ref{parabola})] which are convenient for order-of-magnitude estimates. The main
qualitative difference from the rigorous calculations is that the approximate results become independent of $\kappa$ in the
limit of small $\tilde{d}_{\rm eff}$ (compare with Fig.~\ref{kcrfig}).

\section{Summary and Conclusion}

In this paper we discussed effects of {\it strong} wake-induced coupling between dust-lattice wave modes in 2D plasma
crystals. So far, the mode coupling occurring due to nonreciprocal particle-wake interactions (and resulting in the {\it
mode-coupling instability}, the major plasma-specific mechanism of melting of 2D crystals) has been systematically studied
only in the {\it weak} regime, when the effective dipole moment of wake is assumed to be sufficiently small \cite{Zhdanov09,Couedel10,Roecker12,Roecker12eff}. In this
case, the longitudinal in-plane and out-of-plane modes (participating in the coupling) were shown to be strongly modified
only in a small vicinity of their crossing, where they form the hybrid mode.

In fact, the regime of strong mode coupling has been already achieved in several experiments \cite{Ivlev03,Couedel10,Liu10corr}, and therefore careful
analysis of its implications is absolutely necessary for further studies of 2D plasma crystals. Below we summarize the most
notable features of the strong mode coupling:
\begin{enumerate}[(i)]
\item{The two coupled wave modes corresponding to the in-plane and out-of-plane motion are significantly modified in a
    broad range of wave numbers -- the branches become ``attracted'' to each other even before they merge and form the
    hybrid modes. The modified modes are called, respectively, ``lower'' and ``upper'', to distinguish from the
    weak-coupling regime.}
\item{Unlike the weak-coupling regime, the polarization of the two modes becomes essentially elliptical before the
    hybridization: The (originally longitudinal) lower mode can have a significant transverse component, the (originally
    transverse) upper mode has a longitudinal component.}
\item{The polarization of the hybrid modes is circular at the merging ends, and is (weakly) elliptical in the middle of
    the hybrid branch. The polarization ellipses in this case are rotated by $45^{\circ}$ with respect to the
    polarization of the separate modes, and therefore the hybrid mode spectral density has equal longitudinal and
    transverse components.}
\item{The individual particle trajectories for the separate modes are localized within ellipsoidal ``orbitals'' with the
    mutually orthogonal axes. Upon the hybridization, the orbital representing the (growing) upper mode is significantly
    increased and rotated by an angle notably larger than $45^{\circ}$.}
\item{The location of the hybrid mode can be significantly shifted towards smaller wave numbers, away from the border of
    the Brioullin zone (where the hybrid mode is formed for a weak coupling). The relative magnitude of the effect can
    be as high as $\simeq15\%$.}
\end{enumerate}

It would be highly desirable to perform accurate experimental measurements of the individual particle trajectories and wave
fluctuation spectra (performed with increased signal-to-noise ratio). The analysis of such trajectories and spectra would
make it possible to reveal (at least) some of the above-mentioned features. The theory presented in this paper provides
direct relation between the magnitude of the strong-mode-coupling effects (such as the wave polarization, distribution of
the fluctuation intensity, shift of the hybrid mode) and the principal characteristics of the plasma wakes.

\section*{Acknowledgments}

We appreciate funding from the European Research Council under the European Union's Seventh Framework Programme
(FP7/2007-2013)/ERC Grant agreement 267499.

\begin{appendix}

\section{Elements of the dynamical matrix $\textsf{D}_{k}$}\label{elements}

We introduce utility functions
\begin{eqnarray}
	h_1(\kappa) = e^{-\kappa}(\kappa^{-1}+2\kappa^{-2}+2\kappa^{-3}), \nonumber \\
	h_2(\kappa) = e^{-\kappa}(\kappa^{-2}+\kappa^{-3}), \nonumber \\
	h_3(\kappa) = h_1(\kappa)+h_2(\kappa), \nonumber
\end{eqnarray}
to be used for expressions of the elements of $\textsf{D}_{k}$ [Eq.~(\ref{eigen})]. For the 1D string model with
next-neighbor interactions \cite{Ivlev01} we have:
\begin{widetext}
\begin{eqnarray}
	\Omega^2_{\rm h} = \frac4{1-\tilde{q}} \left(h_1(\kappa) + \tilde{q} h_2(\tilde{\kappa})
-\frac{\tilde{q}}{1+\tilde{\delta}^2} h_3(\tilde{\kappa}) \right) \sin^2 \frac12 k, \nonumber \\
	\Omega^2_{\rm v} = \Omega^2_{\rm conf} - \frac4{1-\tilde{q}} \left(h_2(\kappa) - \tilde{q} h_2(\tilde{\kappa})
+\frac{\tilde{q}\tilde{\delta}^2}{1+\tilde{\delta}^2} h_3(\tilde{\kappa}) \right) \sin^2 \frac12 k, \label{entr1D} \\
	\sigma = 2 \frac{\tilde{q} \tilde{\delta}}{(1-\tilde{q})(1+\tilde{\delta}^2)} h_3(\tilde{\kappa})\sin k. \nonumber
\end{eqnarray}
where $\tilde{\kappa} = \kappa \sqrt{1+\tilde{\delta}^2}$.

The rigorous results for 2D monolayers are obtained by taking into account interactions with all neighbors
\cite{Zhdanov09,Couedel11}. For the wave vector ${\bf k} = k {\bf \hat{e}}_x$, the elements of $\textsf{D}_{k}$ are:
\begin{eqnarray}
	\Omega^2_{\rm h} = \frac2{1-\tilde{q}} \sum_{m,n} \left( \frac{s^2_x}{s^2}h_1(\kappa s) - \frac{s^2_y}{s^2}h_2(\kappa s)
+\tilde{q}h_2(\kappa s_{\delta})-\tilde{q}\frac{s^2_x}{s^2_{\delta}}h_3(\kappa s_{\delta})\right)\sin^2\frac12 k s_x, \nonumber\\
	\Omega^2_{\rm v} = \Omega^2_{\rm conf} - \frac2{1-\tilde{q}} \sum_{m,n} \left(h_2(\kappa s) - \tilde{q}h_2(\kappa s_{\delta})
+\tilde{q}\frac{\tilde{\delta}^2}{s^2_{\delta}}h_3(\kappa s_{\delta})\right) \sin^2 \frac12 k s_x, \label{entr2D} \\
	\sigma=\frac{\tilde{q}\tilde{\delta}}{1-\tilde{q}}\sum_{m,n}\frac{s_{x}}{s^2_{\delta}}h_3(\kappa s_{\delta})\sin k s_x,
\nonumber
\end{eqnarray}
where $s_x = \sqrt{3}m/2$, $s_y = m/2+n$, $s=\sqrt{m^2+n^2+mn}$, $s_{\delta} = \sqrt{s^2+\tilde{\delta}^2}$, the summation
is over all integers except $(0,0)$.
\end{widetext}

\section{Shift of the critical point [Eq.~(\ref{shifts})]}\label{shifteq}

The critical wave number $k_{\rm cr}$ and critical confinement frequency $\Omega_{\rm cr}$ for the onset of hybridization
are determined from the following conditions on the reduced coupling parameter: $p(k_{\rm cr},\Omega_{\rm cr}) = 1$ and
$\partial p/\partial k|_{k_{\rm cr},\Omega_{\rm cr}} = 0$.

For the 1D string model we use Eq.~(\ref{entr1D}), which yields:
\begin{equation}\label{cr1D}
	\Delta k_{\rm cr} = \arctan{\mathcal D}, \quad
	\frac{\Omega^2_{\rm cr}}{\Omega^2_{\rm cr,0}} = \frac12\left(1+\sqrt{1+{\mathcal D}^2}\right),
\end{equation}
where
\begin{equation*}
    {\mathcal D}=\frac{2\tilde q\tilde\delta h_3(\tilde\kappa)}{(1+\tilde\delta^2)h_3(\kappa)-\tilde q(1-\tilde\delta^2)
    h_3(\tilde\kappa)}.
\end{equation*}
For small $\tilde\delta$ we get ${\mathcal D} \to 2\tilde q\tilde\delta/(1-\tilde q)\equiv2\tilde d_{\rm eff}$. Thus, in the
weak-coupling regime we get a simple universal relation between $\Delta k_{\rm cr}$ and $\Delta\Omega_{\rm cr}$,
\begin{equation}\label{parabola}
	\frac{\Delta \Omega_{\rm cr}}{\Omega_{\rm cr,0}} \simeq \frac18 \left(\Delta k_{\rm cr} \right)^2
\simeq \frac12\tilde d_{\rm eff}^2.
\end{equation}

For the rigorous 2D model we employ Eq.~(\ref{entr2D}) and obtain the following equations for $k_{\rm cr}$ and $\Omega_{\rm
cr}$:
\begin{equation}\label{cr2D}
\begin{array}{r}
	{\displaystyle\sum_{m,n} s_x\big({\mathcal A}\cos k_{\rm cr}s_x + {\mathcal B}\sin k_{\rm cr}s_x\big) = 0,} \vspace{0.2cm}\\
    {\displaystyle\frac{\Omega^2_{\rm cr}}{\Omega^2_{\rm cr,0}} =\frac{\sum_{m,n}\left[{\mathcal A}\sin k_{\rm cr}s_x
    +{\mathcal B}\left(1-\cos k_{\rm cr}s_x\right)\right]}{\sum_{m,n}{\mathcal B}\left[1-(-1)^m\right]},}
\end{array}
\end{equation}
where
\begin{eqnarray}
	{\mathcal A}(\tilde{q}, \tilde{\delta}, \kappa) = 4\tilde{q} \tilde{\delta} \frac{s_x}{s^2_{\delta}}h_3(\kappa s_{\delta}),
\nonumber\\
	{\mathcal B}(\tilde{q}, \tilde{\delta}, \kappa) = 2\frac{s^2_x}{s^2}h_3(\kappa s)-2\tilde{q} \frac{s_x^2-\tilde{\delta}^2}
{s^2_{\delta}}h_3(\kappa s_{\delta}).\nonumber
\end{eqnarray}
The first equation~(\ref{cr2D}) yields $k_{\rm cr}$, which should then be substituted to the second equation to obtain
$\Omega_{\rm cr}$.

\end{appendix}

\end{document}